\documentstyle[twoside,fleqn,npb,epsfig,epsf,feynmp,here]{article}
\def\cK{{\cal K}} 
\def \be {\begin{equation}}
\def \ee {\end{equation}}
\def \bea {\begin{eqnarray}}
\def \ea {\end{eqnarray}}
\def \ep {\varepsilon}
\def \ga {\gamma}
\def \la {\lambda}
\def \no {\nonumber}
\def \dd {{\rm d}}
\def \ps {p\hspace{-0.43em}/}
\def \mps {p\hspace{-0.45em}/}
\def \ns {n\hspace{-0.51em}/}
\def \mns {n\hspace{-0.53em}/}
\def \pa {\partial}
\def \c {\hspace{-0.2em} \cdot}
\title{Electroweak Sudakov logarithms in the Coulomb gauge%
       \hspace*{0.3ex}\thanks{Talk presented by W.~Beenakker at {\it Loops and 
       Legs in Quantum Field Theory}, Bastei, Germany, April 2000}}

\author{W. Beenakker%
        \hspace*{0.3ex}$\vphantom{B}^{\rm a}$\thanks{Supported by a PPARC 
                                                     Research Fellowship}
        and
        A. Werthenbach%
        \hspace*{0.3ex}\address{Department of Physics, University of Durham, 
                                Durham DH1 3LE, U.K.}%
        \thanks{Supported by a DAAD Doktorandenstipendium (HSP III)}}
\begin{document}

\begin{abstract}
We describe a formalism for calculating electroweak Sudakov logarithms in the
Coulomb gauge. This formalism is applicable to arbitrary electroweak processes.
For illustration we focus on the specific reactions $\,e^+e^- \to f \bar{f}\,$ 
and $\,e^+e^- \to W_T^+W_T^-,\,W_L^+W_L^-$, which contain all the salient 
details of dealing with the various types of particles. We discuss an explicit 
two-loop calculation and have a critical look at the (non-)exponentiation and 
factorisation properties of the Sudakov logarithms in the Standard Model. 
\end{abstract}

\maketitle

\section{Introduction} 
At the next generation of linear $e^+e^-$ colliders TeV-scale center-of-mass 
energies will be reached \cite{LCreport}. At these energies the effects 
arising from weak corrections are expected to be of the order 
of $10\%$ or more~\cite{size,wwhea}, i.e.\ just as large as the well-known
electromagnetic corrections. In order not to jeopardize any of the 
high-precision studies at these high-energy colliders, it is therefore 
indispensable to improve the theoretical understanding of the radiative 
corrections in the weak sector of the Standard Model (SM). In particular this 
will involve a careful analysis of effects beyond first order in the 
perturbative expansion in the (electromagnetic) coupling $\alpha=e^2/(4\pi)$.

The dominant source of radiative corrections at TeV-scale energies 
is given by the so-called Sudakov logarithms 
$\propto \alpha^n\log^{2n}(M^2/s)$, involving particle masses $M$ well 
below the collider energy $\sqrt{s}$. These corrections have a unique origin, 
being related to collinear-soft singularities~\cite{sudakov}.
In a recent study we have investigated these Sudakov effects at two-loop level
in the process $e^+e^- \to f\bar{f}$ \cite{nonexp}, finding disagreement with
three earlier studies \cite{kuehn1}--\cite{melles}. Here we give a 
description of our formalism, which is based on the Coulomb gauge, and extend
it to reactions with transverse and longitudinal (massive) gauge bosons in 
the final state. Especially the treatment of the longitudinal gauge bosons 
requires some special attention. In order to cover all the relevant features
and subtleties of our method, it is sufficient to restrict the discussion to
the virtual corrections. In fact, since the Sudakov logarithms originate from 
the exchange of soft, effectively on-shell gauge bosons, many of the features 
derived for these virtual corrections are intimately related to properties of 
the corresponding real-gauge-boson emission processes.

\section{Electroweak Sudakov logarithms in the Coulomb gauge}

In order to facilitate the calculation of the one- and two-loop Sudakov 
logarithms, we work in the Coulomb gauge for both massless and massive gauge
bosons. The power of this gauge choice lies in the fact that the gauge-boson
propagators become effectively transverse:
\bea
  P^{\mu\nu}(k) \!\!\!\!&=&\!\!\!\! -\,i\,\frac{\vec{k}^{\,2}\,g^{\mu\nu}
        + k^{\mu}k^{\nu} - k^0\,( k^{\mu}n^{\nu}+n^{\mu}k^{\nu} )}
        {\vec{k}^{\,2}\,(k^2-M^2+i \epsilon)} \no\\[1mm]
                \!\!\!\!&=&\!\!\!\! \frac{-\,i}{k^2-M^2+i \epsilon}
       \Bigl[ Q^{\mu\nu}(k) - \frac{k^2}{\vec{k}^{\,2}}\,n^{\mu}n^{\nu}
       \Bigr].     
\ea
Here $k$ and $M$ are the momentum and mass of the gauge boson, and $n$ is the 
unit vector in the time direction, which enters by virtue of using the Coulomb
gauge. The tensor
\be
  Q_{\mu\nu}(k) = - \sum\limits_{\la=T}\ep_{\mu}(k,\la)\,\ep_{\nu}^*(k,\la)
\ee
is the polarization sum for the transverse helicity states. Therefore the 
gauge bosons are effectively transverse if $\,k^2 \ll \vec{k}^{\,2}$, which
is the case for collinear gauge-boson emission at high energies 
($k^2 \propto M^2$ and $\vec{k}^{\,2} \approx k_0^2 \gg M^2$). As a result
of the effective transversality, the virtual Sudakov logarithms originating 
from vertex, box etc.\ corrections are suppressed (provided all kinematical
invariants are of the same order as the CM energy squared). Hence, all virtual 
Sudakov logarithms are contained exclusively in the self-energies of the 
external on-shell particles~\cite{taylor,coulomb} or the self-energies of any 
intermediate particle that happens to be effectively on-shell. The latter is, 
for instance, needed for the production of near-resonance unstable particles. 
The elegance of this method lies in its universal nature. Once all 
self-energies to all on-shell/on-resonance SM particles have been calculated, 
the prediction of the Sudakov form factor for an {\it arbitrary} electroweak 
process becomes trivial. The relevant self-energies for the calculation of the
Sudakov logarithms involve the exchange of collinear-soft gauge bosons. 
The collinear-soft exchange of fermions, scalars and ghosts leads to suppressed
contributions, since the propagators of these particles do not have the 
required pole structure.

\subsection{The external wave-function factors}

The calculation of the external wave-function factors in the Coulomb gauge is
rather straightforward for the fermions. For massive gauge bosons, however, 
the mixing with the corresponding component of the Higgs doublet introduces an 
additional complication. For instance, consider the $W$ boson and the
would-be Goldstone boson $\phi$. For a proper description of the on-shell
$W$ bosons we have to define the asymptotic $W^{\rm as}$ field in terms of the 
interacting $W$ and $\phi$ fields: 
\bea
  W_{\mu}^{\pm,\,\rm as}(x) \!\!&=&\!\! 
          Z^{-\frac{1}{2}}_{_W}\,W_{\mu}^{\pm}(x)
          \,\pm\, i\,\delta Z_{\phi}\,\frac{\pa_{\mu}\phi^{\pm}(x)}{M_W} 
          \no\\[1mm] 
                            \!\!& &\!\!\hspace*{-10ex}{}
          {}+ \delta Z_n\, n_{\mu}n\,\c W^{\pm}(x)  
          + \delta Z_{\pa}\,\frac{\pa_{\mu}\,\pa\,\c W^{\pm}(x)}{M_W^2},
\ea
in such a way that the lowest-order (`free-field') propagators are retrieved 
for $W^{\rm as}$. This fixes the renormalization factors $Z$ and $\delta Z$
in terms of the self-energies of the interacting fields \cite{coulomb}.

For transverse polarization states (T) the mixing with the $\phi$ field 
vanishes and the asympotic state effectively reduces to
\be
  W_{\mu}^{\pm,\,\rm as}(x) \stackrel{T}{\to} 
  Z^{-\frac{1}{2}}_{_W}\,W_{\mu}^{\pm}(x),
\ee
since $\,\ep_T(k)\,\c k = \ep_T(k)\,\c n = 0$. The transverse wave-function 
factor $Z_{_W}$ is obtained from the purely transverse $Q_{\mu\nu}$ part of the
Dyson-resummed gauge-boson self-energy \cite{coulomb}. The contribution of 
Sudakov logarithms now simply amounts to multiplying each external transverse 
gauge-boson line of the matrix element by the factor $\,Z^{\frac{1}{2}}_{_W}$. 

For longitudinal polarization states (L) we do have to deal with the 
$W$--$\phi$ mixing. In that case the relevant part of the asymptotic state is
\be
  W_{\mu}^{\pm,\,\rm as}(x) \stackrel{L}{\to} 
          Z^{-\frac{1}{2}}_{_W}\,W_{\mu}^{\pm}(x)
          + \delta Z_n\, n_{\mu}n\,\c W^{\pm}(x),
\ee
since $\,\ep_L(k)\,\c k = 0\,$ and $\,\ep_L(k)\,\c n \approx k_0/M_W$.
In this case only the $n_{\mu}n_{\nu}$ part of the gauge-boson self-energy and
the $n_{\mu}$ part of the $W$--$\phi$ mixing self-energy contribute. In the
Sudakov limit these self-energies are related \cite{coulomb}, resulting in the
following identity for external longitudinal $W$ bosons:
\bea
\label{ET}
\hspace*{-3mm} & &\begin{picture}(85,20)(8,38)
     \begin{fmffile}{gw1}
     \begin{fmfgraph*}(80,80) \fmfpen{thin} \fmfleft{i1} \fmfright{o1}
     \fmf{phantom}{i1,v1,v2,o1} 
     \fmf{plain,left,tension=1.5}{i1,v1} \fmf{plain,right,tension=1.5}{i1,v1}
     \fmf{boson,tension=4}{v1,v2} \fmfblob{.15w}{v2}
     \fmf{double,tension=4}{v2,o1}
     \fmfdot{o1}
     \fmfv{label={$\textstyle \nu$},l.a=-90,l.d=7}{o1}
     \fmfv{label={$\textstyle W^{\mathrm as}$},l.a=15,l.d=45}{v1}
     \fmfv{label={$\textstyle W$},l.a=135,l.d=12}{v2}
     \fmfv{label={$\textstyle \mu$},l.a=-15,l.d=40}{i1}
     \end{fmfgraph*}
     \end{fmffile}
     \end{picture}
     \hspace*{2ex} i\,(k^2-M_W^2)\,\ep^{\nu}_{L}(k)
     \\[8mm]
\hspace*{-3mm} &+&\begin{picture}(85,20)(8,38)
     \begin{fmffile}{gphi1}
     \begin{fmfgraph*}(80,80) \fmfpen{thin} \fmfleft{i1} \fmfright{o1}
     \fmf{phantom}{i1,v1,v2,o1} 
     \fmf{plain,left,tension=1.5}{i1,v1} \fmf{plain,right,tension=1.5}{i1,v1}
     \fmf{dashes,tension=4}{v1,v2} \fmfblob{.15w}{v2}
     \fmf{double,tension=4}{v2,o1}
     \fmfdot{o1}
     \fmfv{label={$\textstyle \nu$},l.a=-90,l.d=7}{o1}
     \fmfv{label={$\textstyle W^{\mathrm as}$},l.a=15,l.d=45}{v1}
     \fmfv{label={$\textstyle \phi$},l.a=138,l.d=17}{v2}
     \end{fmfgraph*}
     \end{fmffile} 
     \end{picture}  
     \hspace*{2ex} i\,(k^2-M_W^2)\,\ep^{\nu}_{L}(k)
     \no\\[8mm]
\hspace*{-3mm} &\approx& \begin{picture}(47,20)(6,30)
      \begin{fmffile}{gphi2}
      \begin{fmfgraph*}(65,65) \fmfpen{thin} \fmfleft{i1} \fmfright{o1}
      \fmf{phantom}{i1,v1,v2,o1}
      \fmf{plain,left,tension=1.5}{i1,v1} \fmf{plain,right,tension=1.5}{i1,v1}
      \fmf{phantom,tension=4}{v1,v2} \fmf{phantom,tension=4}{v2,o1}
      \fmf{dashes}{v1,v2}
      \fmfv{label={$\textstyle \phi$},l.a=120,l.d=10}{v2}
      \end{fmfgraph*} 
      \end{fmffile}
      \end{picture}
      \hspace*{0.5ex} Z^{\frac{1}{2}}_{_L} 
  \ -\ 
      \begin{picture}(47,20)(4,30)
      \begin{fmffile}{gw2}
      \begin{fmfgraph*}(65,65) \fmfpen{thin} \fmfleft{i1} \fmfright{o1}
      \fmf{phantom}{i1,v1,v2,o1}
      \fmf{plain,left,tension=1.5}{i1,v1} \fmf{plain,right,tension=1.5}{i1,v1}
      \fmf{phantom,tension=4}{v1,v2} \fmf{phantom,tension=4}{v2,o1}
      \fmf{boson}{v1,v2}
      \fmfv{label={$\textstyle W$},l.a=120,l.d=10}{v2}
      \fmfv{label={$\textstyle \mu$},l.a=-160,l.d=35}{o1}
      \end{fmfgraph*} 
      \end{fmffile} 
      \end{picture}
      \hspace*{1ex}\frac{M_W}{k_0}\,Z^{\frac{1}{2}}_{_L}\,n^{\mu}. \no 
\ea \vspace*{0.1cm}

\noindent Here the last two diagrams are amputated at the external line. 
The longitudinal wave-function factor $Z_{_L}$, defined as 
\be
  Z^{-\frac{1}{2}}_{_L} \equiv Z^{-\frac{1}{2}}_{_W} + \delta Z_n,
\ee 
is obtained from the self-energy of the scalar $\phi$ field.
The identity in Eq.~(\ref{ET}) is in fact simply the Equivalence Theorem, 
stating that a non-vanishing matrix element for longitudinal $W$ bosons at high
energies is equivalent to the corresponding matrix element with the $W$ bosons
replaced by the would-be Goldstone bosons $\phi$. Hence, at high energies the 
would-be Goldstone bosons effectively become physical Goldstone bosons, at the
expense of the longitudinal degrees of freedom of the massive gauge bosons.
This is exactly what one would expect if the SM were to behave like an unbroken
theory at high energies.

\section{The Sudakov logarithms at one-loop}

As an example we sketch the calculation of the one-loop Sudakov logarithms in 
the process $e^+e^- \to f \bar{f}$. Consider to this end the fermionic 
one-loop self-energy $\Sigma^{\,(1)}$, describing the emission of a 
gauge boson $V_1$ with loop-momentum $k_1$ and mass $M_1$ from a fermion $f$ 
with momentum $p$ and mass $m_f$:
\begin{displaymath}
  -i\,\Sigma^{\,(1)}(p,n) \,= 
  \begin{picture}(25,35)(4,67)
    \begin{fmffile}{self}
    \begin{fmfgraph*}(140,140) \fmfpen{thin} \fmfleft{i1} \fmfright{o1}
      \fmf{fermion,tension=0.7,label=$f(p)$,l.side=right,width=0.6}{i1,v1} 
      \fmf{fermion,tension=0.5,label=$f_1(p\!-\!k_1)$,l.side=right,width=0.6}
          {v1,v3} 
      \fmf{fermion,tension=0.7,label=$f(p)$,l.side=right,width=0.6}{v3,o1} 
      \fmfdot{v1} \fmfdot{v3} 
      \fmf{boson,left,tension=-1,label=$V_1(k_1)$,l.side=left,width=0.6}{v1,v3}
      \fmf{phantom,left}{v1,v3}
      \fmffreeze
      \fmf{phantom_arrow}{v1,v3}
    \end{fmfgraph*}
    \end{fmffile}
  \end{picture}
\end{displaymath} \vspace*{0.5cm}

\noindent In the high-energy limit the fermion mass 
in the numerator of the fermion propagator can be neglected. The self-energy 
$\Sigma^{\,(1)}$ then contains an odd number of $\ga$-matrices, leading to 
the following natural decomposition in terms of the two possible structures 
$\ps$ and $\ns$:
\bea
  \Sigma^{\,(1)}(p,n) \!\!&\approx&\!\! 
      e^2\,\Gamma_{\!\!ff_1\!V_1}^{\,^{\scriptstyle 2}}\,
      \Bigl[ \,\mns\:\frac{p^2}{n\,\c p}\,\Sigma_n^{\,(1)}(n\,\c p,p^2)
      \no \\[1mm]
                      \!\!&       &\!\! 
      \hphantom{e^2\,\Gamma_{\!\!ff_1\!V_1}^{\,^{\scriptstyle 2}}a}           
      {}+ \mps\,\Sigma_p^{\,(1)}(n\,\c p,p^2)
      \,\Bigr].
\ea
The coupling factor $\Gamma_{\!\!ff_1\!V_1}$ is defined according to
\be
   \Gamma_{\!\!ff_1\!V_1} = V_{\!\!ff_1\!V_1} - \ga_5\,A_{\!ff_1\!V_1},
\ee
where $V_{\!\!ff_1\!V_1}$ and $A_{\!ff_1\!V_1}$ are the vector and axial-vector
couplings of the fermion $f$ to the exchanged gauge boson $V_1$. 

The corresponding one-loop contribution to the external wave-function factor 
$Z_f = 1 + \delta Z_f$ can be obtained by means of the projection~\cite{nonexp}
\bea
  \delta Z_f^{\,(1)} \!\!\!\!&=      &\!\! \frac{1}{2\,p_0}\,\bar{u}_f(p)\,
           \biggl\{ \frac{\pa}{\pa p_0}\,\Sigma^{\,(1)}(p,n)
           \biggr\}\,u_f(p) \\[1mm]
                     \!\!\!\!&\approx& \!\! -\int\!\frac{\dd^4 k_1}{(2\,\pi)^4}
           \,\frac{4\,e^2\,\Gamma_{\!\!ff_1\!V_1}^{\,^{\scriptstyle 2}}\,
                 p_{\mu}\,p_{\nu}\,P^{\mu\nu}(k_1,n)}
                {[(p-k_1)^2-m_{f_1}^2+i\epsilon]\,^2}~. \no
\ea
In the last step we have exploited the fact that only collinear-soft 
gauge-boson momenta give rise to the Sudakov logarithms. Bearing in mind the 
effective transversality of the gauge-boson propagator $P^{\mu\nu}$ in that 
limit, the final expression exhibits the usual eikonal behaviour expected for 
the exchange of a soft gauge boson. 

Having two canonical momenta at our disposal, i.e.\ $p$ and
$n$, we define the following Sudakov parametrisation of the gauge-boson 
loop-momentum $k_1$:
\bea
  k_1       \!\!&=&\!\! v_1\,q + u_1\,\bar{q} + k_{1_\bot}, 
\ea
with 
\bea
  p^{\mu} \equiv (E,\beta_f E,0,0), 
  & & q^\mu = (E,E,0,0), \no\\[1mm]                
  \bar{q}^\mu = (E,- E,0,0),\ 
  & & k_{1_\bot}^{\mu} = (0,0,\vec{k}_{1_\bot}),
\ea
where $\,\beta_f = \sqrt{1-m_f^2/E^2}\,$ and $\,E \equiv \sqrt{s}/2\,$ are the 
velocity and energy of the external fermion~$f$.

The $v_1$-integration is restricted to the interval $\,0\le v_1 \le 1$, as 
a result of the requirement of having poles in both hemispheres of the complex 
$u_1$-plane. The residue is then taken in the lower hemisphere in the pole of
the gauge-boson propagator: $s\,v_1\,u_1^{\rm res} = \vec{k}_{1_\bot}^{\,2} 
+ M_1^2 \equiv s\,v_1\,y_1$. Finally, $\vec{k}_{1_\bot}^{\,2}$ is substituted 
by $y_1$, with the condition $\,\vec{k}_{1_\bot}^{\,2} \ge 0\,$ translating 
into $\,v_1\,y_1 \ge M_1^2/s$. The one-loop Sudakov contribution to 
$\delta Z_f$ now reads 
\be
\label{kernel}
  \delta Z_f^{\,(1)} \!\approx\, -\,\frac{\alpha}{\pi}\,
     \Gamma_{\!\!ff_1\!V_1}^{\,^{\scriptstyle 2}}\!
     \int_0^1 \!\frac{\dd y_1}{y_1}\!
     \int_{y_1}^1 \!\frac{\dd z_1}{z_1}\ \cK(y_1,z_1),
\ee
with the integration kernel $\,\cK(y_1,z_1)\,$ given by 
\be
  \cK(y_1,z_1) \,=\, \Theta\Bigl( y_1 z_1-\frac{M_1^2}{s} \Bigr)
                     \Theta\Bigl (y_1-\frac{m_f^2}{s}\,z_1\Bigr).
\ee
Here we introduced the energy variable $\,z_1 = v_1 + y_1\,$ and made use of 
the fact that only collinear-soft gauge-boson momenta are responsible for the 
quadratic large-logarithmic effects: $y_1,\,z_1 \ll 1$. As a result, the 
gauge boson inside the loop is effectively on-shell and transversely 
polarized. The $\theta$-function containing the fermion mass $m_f$ is needed 
for the exchange of photons only, regulating the collinear singularity at 
$y_1=0$. For the exchange of a massive gauge boson the mass $M_1$ will be the 
dominant collinear as well as infrared regulator. The final step of 
calculating the double (logarithmic) integral in Eq.~(\ref{kernel}) is now
trivial.

The calculation of the Sudakov logarithms for other types of external particles
proceeds in a similar way, using appropriate projection methods to bring the
correction to the wave-function factor in an eikonal form \cite{coulomb}.

Upon summation over the allowed gauge-boson exchanges, one obtains the 
following expression for the full one-loop Sudakov correction to the external 
wave-function factor for an arbitrary particle:
\bea
\label{delta1}
  \delta Z^{\,(1)} \!\!&=&\!\! \left[ \frac{C_2(R)}{\sin^2\theta_{\rm w}} 
                        + \left( \frac{Y}{2\cos\theta_{\rm w}} \right)^{\!2}\, 
                        \right]\! {\rm L}(M,M) \no\\[1mm]
                   \!\!& &\!\! {}+ Q^2\,\bigg[ {\rm L_{\ga}}(\la,m)
                        - {\rm L}(M,M) \bigg],
\ea
with $\theta_{\rm w}$ the weak mixing angle and
\bea
\label{L}
 {\rm L_{\ga}}(\la,M_1)\ \!\!\!&=&\!\!\! -\,\frac{\alpha}{4\,\pi}\! 
          \left[ \log^{\, 2}\!\left( \frac{\la^2}{s} \right) 
           \!-\! \log^{\, 2}\!\left( \frac{\la^2}{M_1^2} \right) 
          \right], \no\\[1mm]
 {\rm L}(M_1,M_2)       \!\!\!&=&\!\!\! -\,\frac{\alpha}{4\,\pi}
          \log\left( \frac{M_1^2}{s} \right) 
          \log\left( \frac{M_2^2}{s} \right). 
\ea
Here $\,m,\,eQ,\,Y\,$ and $\,C_2(R)\,$ are the mass, charge, hypercharge and
$SU(2)$ Casimir operator of the external particle. So, $C_2(R) = C_F = 3/4\,$
for the fermions and longitudinal gauge bosons (read: Goldstone bosons), which
are in the fundamental representation, and $C_2(R) = C_A = 2\,$ for transverse
gauge bosons, which are in the adjoint representation. Finally, $\la$ is the 
fictitious (infinitesimally small) mass of the photon needed for regularizing 
the infrared singularity at $z_1=0$. For the sake of calculating the leading 
Sudakov logarithms, the masses of the $W$ and $Z$ bosons can be represented by 
one generic mass scale $M$. Note that the terms proportional to $Q^2$ in 
Eq.~(\ref{delta1}) are the result of the mass gap between the photon and the 
weak bosons. 

We have applied these one-loop Sudakov correction factors to the reactions 
$\,e^+e^- \to W_T^+W_T^-,\,W_L^+W_L^-\,$ and found perfect agreement with
the high-energy approximation in Ref.~\cite{wwhea}, which confirms the
afore-mentioned differences between transverse and longitudinal degrees of
freedom.

\section{Exponentiation revisited}

\begin{figure*}

\vspace*{0.2cm}
\hspace*{10.4cm}{\epsfysize=1.65cm{\epsffile{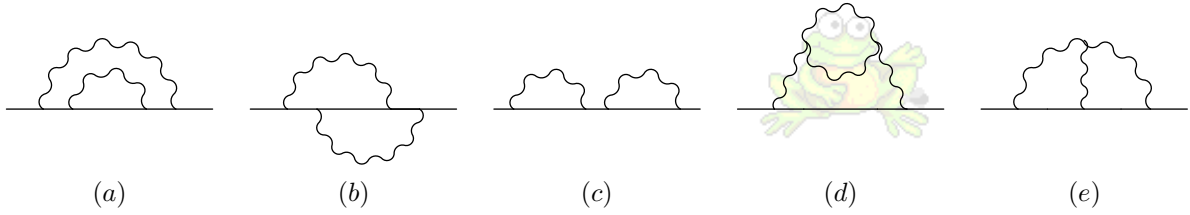}}}
\vspace*{-1.3cm}

\begin{fmffile}{2l}
\hspace*{0.2cm}
\begin{fmfgraph*}(78,52) \fmfpen{thin} \fmfleft{i1} \fmfright{o1}
\fmf{phantom,tension=1}{i1,v1,v2,v5,v3,v4,o1}
\fmf{plain,tension=1,l.side=right,width=0.6}{i1,v1} 
\fmf{plain,tension=1.5,l.side=right,width=0.6}{v1,v2}
\fmf{plain,tension=0.5,l.side=right,width=0.6}{v2,v3}
\fmf{plain,tension=1.5,l.side=right,width=0.6}{v3,v4}
\fmf{plain,tension=1,l.side=right,width=0.6}{v4,o1}
\fmffreeze 
\fmf{boson,left,tension=-1,l.side=left,l.dist=4,width=0.6}{v1,v4}
\fmf{boson,left,tension=-2,l.side=left,l.dist=2,width=0.6}{v2,v3}
\fmfv{label=$(a)$,l.dist=27,l.a=-90}{v5}
\end{fmfgraph*}
\hspace{0.3cm}
\begin{fmfgraph*}(78,52) \fmfpen{thin} \fmfleft{i1} \fmfright{o1}
\fmf{phantom,tension=1}{i1,v1,v2,v5,v3,v4,o1}
\fmf{plain,tension=1,l.side=right,width=0.6}{i1,v1} 
\fmf{plain,tension=1.5,l.side=right,width=0.6}{v1,v2}
\fmf{plain,tension=0.5,width=0.6}{v2,v3}
\fmf{plain,tension=0.1,width=0.6}{v3,v4}
\fmf{plain,tension=1.5,l.side=left,width=0.6}{v3,v4}
\fmf{plain,tension=1,l.side=right,width=0.6}{v4,o1}
\fmffreeze 
\fmf{boson,left,tension=-1.5,l.side=left,l.dist=4,width=0.6}{v1,v3}
\fmf{boson,right,tension=-1.5,l.side=right,l.dist=4,width=0.6}{v2,v4}
\fmfv{label=$(b)$,l.dist=27,l.a=-90}{v5}
\end{fmfgraph*}
\hspace{0.3cm}
\begin{fmfgraph*}(78,52) \fmfpen{thin} \fmfleft{i1} \fmfright{o1}
\fmf{phantom,tension=1}{i1,v1,v2,v5,v3,v4,o1}
\fmf{plain,tension=3,l.side=right,width=0.6}{i1,v1} 
\fmf{plain,tension=0.1,l.side=right,width=0.6}{v1,v2}
\fmf{plain,tension=3,l.side=right,width=0.6}{v2,v3}
\fmf{plain,tension=0.1,l.side=right,width=0.6}{v3,v4}
\fmf{plain,tension=3,l.side=right,width=0.6}{v4,o1}
\fmffreeze 
\fmf{boson,left,tension=1,l.side=left,l.dist=4,width=0.6}{v1,v2}
\fmf{boson,left,tension=1,l.side=left,l.dist=4,width=0.6}{v3,v4}
\fmfv{label=$(c)$,l.dist=27,l.a=-90}{v5}
\end{fmfgraph*}
\hspace{0.3cm}
\begin{fmfgraph*}(78,52) \fmfpen{thin} \fmfleft{i1} \fmfright{o1}
\fmf{phantom,tension=1}{i1,v1,v2,v5,v3,v4,o1}
\fmf{plain,tension=1,l.side=right,width=0.6}{i1,v1} 
\fmf{plain,tension=1.5,width=0.6}{v1,v2}
\fmf{plain,tension=0.5,l.side=right,width=0.6}{v2,v3}
\fmf{plain,tension=1.5,width=0.6}{v3,v4}
\fmf{plain,tension=1,l.side=right,width=0.6}{v4,o1}
\fmffreeze
\fmftop{v6,v7,v8,v9}
\fmf{boson,tension=-1,l.side=left,l.dist=3,width=0.6}{v1,v7}
\fmf{boson,tension=-1,l.side=left,l.dist=3,width=0.6}{v8,v4}
\fmf{boson,left,tension=2,l.side=left,l.dist=4,width=0.6}{v7,v8}
\fmf{boson,right,tension=2,l.side=left,l.dist=5,width=0.6}{v7,v8}
\fmfv{label=$(d)$,l.dist=27,l.a=-90}{v5}
\end{fmfgraph*}
\hspace{0.3cm}
\begin{fmfgraph*}(78,52) \fmfpen{thin} \fmfleft{i1} \fmfright{o1}
\fmf{phantom,tension=1}{i1,v1,v2,v5,v3,v4,o1}
\fmf{plain,tension=1,l.side=right,width=0.6}{i1,v1} 
\fmf{plain,tension=1.5,width=0.6}{v1,v2}
\fmf{plain,tension=0.5,width=0.6}{v2,v3}
\fmf{plain,tension=1.5,width=0.6}{v3,v4}
\fmf{plain,tension=1,l.side=right,width=0.6}{v4,o1}
\fmffreeze
\fmftop{v6}
\fmf{boson,left,tension=-1,width=0.6}{v1,v4}
\fmf{boson,tension=2,l.side=left,l.dist=3,width=0.6}{v5,v6}
\fmfv{label=$(e)$,l.dist=27,l.a=-90}{v5}
\end{fmfgraph*}
\end{fmffile}
\vspace*{-5mm} 
\caption[]{Generic two-loop fermionic self-energy diagrams contributing to
           the Sudakov logarithms}
\label{twoloop} 
\end{figure*}

In order to discuss the (non-)exponentiation and factorisation properties of 
the Sudakov logarithms in the SM, we now focus on the reaction 
$\,e^+e^- \to f \bar{f}\,$ at two-loop level. At two-loop accuracy one has to 
take into account the five generic sets of diagrams displayed in 
Fig.~\ref{twoloop}.
Various cancellations are going to take place between all these diagrams. 
In unbroken theories like QED and QCD merely the so-called `rainbow' diagrams 
of set (a) survive, and the resummation of the higher-order Sudakov effects 
amounts to an exponentiation of the one-loop corrections (see for instance 
Refs.~\cite{sudakov,taylor,expon,frenkel}\,). The same holds if all 
gauge bosons of the theory would have
a similar mass. The unique feature of the SM is that it is only partially
broken, with the electromagnetic gauge group $U(1)_{\rm em} \neq U(1)_Y$ 
remaining unbroken. As such three of the four gauge bosons will acquire a mass,
whereas the photon remains massless and will interact with the charged massive
gauge bosons ($W^{\pm}$). As a result, the `rainbow' diagrams are not going 
to be the only contributions that survive the gauge cancellations.

The generic two-loop contributions of Sudakov logarithms to 
$\delta Z_f$ can be found in Ref.~\cite{nonexp}. We merely note that 
several of the contributions involve a specific ordering in the energy 
variables $z_i$ [in set (d) and part of set (e)] and/or the angular variables 
$y_i$ [in sets (a),(d) and part of set (e)]. Note also that certain diagrams 
look possible at first sight, but are in fact 
forbidden as a result of the charged current interactions of the $W$ bosons. 
For instance, in set (b) it is not possible to exchange two $W$ bosons without 
reversing the fermion-number flow (given by the direction of the Dirac 
propagator lines). Adding up all possible contributions, we find for the full 
two-loop Sudakov correction factor for right- and left-handed 
fermions/antifermions
\bea
\label{delta2}
 \delta Z^{\,(2)}_{f_L/\bar{f}_R} \!\!\!\!&=&\!\!\!\! \frac{1}{2}
        \left( \delta Z^{\,(1)}_{f_L}\right)^2
      + \left( Q_f^2 - \frac{|Q_f|}{2\,\sin^2\theta_{\rm w}}\right)\Delta_f,
      \no\\[1mm]
 \delta Z^{\,(2)}_{f_R/\bar{f}_L} \!\!\!\!&=&\!\!\!\! \frac{1}{2}
        \left( \delta Z^{\,(1)}_{f_R}\right)^2 +\, Q_f^2\,\Delta_f,
\ea
with
\bea
\label{deltaf}
  \Delta_f = {\rm L}(M,M)\biggl[ \frac{4}{3}\,{\rm L}(M,m_f)
                                 - {\rm L}(M,M) \biggr]
\ea
and ${\rm L}(M_1,M_2)$ as defined in Eq.~(\ref{L}). The expressions for 
transverse and longitudinal gauge bosons have a similar form \cite{coulomb}.

{}From Eq.~(\ref{delta2}) we deduce the main statement of Ref.~\cite{nonexp},
namely that the virtual electroweak two-loop Sudakov correction factor is not 
obtained by a mere exponentiation of the one-loop Sudakov correction factor. 
Based on the explicit two-loop calculation we find non-exponentiating terms, 
originating from the mass gap between the {\it massless} photon and the 
{\it massive} $Z$ boson in the neutral sector of the SM. We have checked that 
these extra terms vanish in leading-logarithmic approximation if {\it all} 
gauge bosons have the same (or roughly the same) mass. From Eq.~(\ref{deltaf})
it is clear that the non-exponentiating terms are genuine quadratic 
large-logarithmic effects. They will not vanish if the fermion mass is of the 
order of the masses of the weak bosons or if the energy is taken to infinity,
since in those cases $\Delta_f \to {\rm L}^2(M,M)/3$. Therefore, as far as the
virtual Sudakov logarithms are concerned, the SM will
never completely behave like an unbroken theory, even not if the energy becomes
arbitrarily large. This is a consequence of the fact that the photon is
massless, i.e.\ $\,m_f/\la \gg \sqrt{s}/M$.

We also note that, in adding up all the contributions, we find that the 
`rainbow' diagrams of set (a) yield the usual exponentiating terms plus an 
extra term similar to the one found in Ref.~\cite{paolo2}, originating from 
the charged-current interactions. Whereas in Ref.~\cite{paolo2} this extra 
term was interpreted as the source of non-exponentiation, we observe that it 
in fact cancels against a specific term originating from the triple gauge-boson
diagrams of set (e). Similar (gauge) cancellations take place between the 
`crossed rainbow' diagrams of set (b), the reducible diagrams of set (c), and 
another part of the triple gauge-boson diagrams of set (e). Finally, most of 
the left-over terms of set (e) get cancelled by the contributions from the
gauge-boson self-energy (`frog') diagrams of set (d). The term proportional to 
$\,|Q_f|\,$ in Eq.~(\ref{delta2}) is the only surviving term of set (e), 
whereas the left--right symmetric term proportional to $\,Q_f^2\,$ originates 
from those diagrams of set (d) that involve neutral gauge bosons in the outer 
loop. The cancellation that usually takes place in unbroken gauge theories is
upset by the fact that the on-shell poles for photons and $Z$ bosons do 
{\it not} coincide, leading to different results for the on-shell residues.
As was pointed out in Ref.~\cite{nonexp}, the energies of both the photon and 
the $Z$-boson are in the weak (soft-energy) domain, $z_1\ge M/\sqrt{s}$, owing 
to the energy ordering. The observed difference is therefore caused entirely 
by the differences in the collinear domain induced by the mass gap. Based on 
this observation we conclude that the statements in Ref.~\cite{paolo2} 
concerning the factorization and exponentiation properties of the Sudakov 
logarithms in the ultrasoft energy regime, 
$\la/\sqrt{s} \le z_1 \le M/\sqrt{s}$, are not contradicted by our analysis.   

Comparing with the other two studies, we can make the following 
remarks. First of all, a treatment of pure weak gauge-boson effects without 
reference to the photonic interactions breaks gauge-invariance, since the 
photon has an explicit $SU(2)$ component. This holds even if the photon is 
treated fully inclusively as in Ref.~\cite{kuehn1}. Such a separation would 
require a very careful definition, for instance in terms of the typical energy 
regimes that govern the Sudakov effects of pure electromagnetic origin 
(ultrasoft energies) and collective electroweak origin (soft energies). 
Second, in contrast to Ref.~\cite{melles} we do not observe the exponentiation 
of the virtual one-loop Sudakov logarithms. In the dispersive method of 
Ref.~\cite{melles} it is assumed that QCD-like diagrammatic cancellations will 
take place, whereas we find that such cancellations can be upset by the fact 
that the on-shell poles for photons and $Z$ bosons do not coincide. The latter 
might also have repercussions on the dispersive method itself, since in the 
diagrams of set (d) with one photon and one $Z$ boson in the outer loop, it is
not possible that both gauge-bosons are effectively on-shell simultaneously.

\section{Conclusions and Outlook} 

We have presented a universal formalism for calculating high-energy electroweak
Sudakov logarithms in the Coulomb gauge. In this special gauge all the relevant
contributions, involving the exchange of collinear-soft gauge bosons, are 
contained in the self-energies of the external on-shell particles. In this
context the treatment of longitudinal gauge bosons requires special care in 
view of their
mixing with the would-be Goldstone bosons. After defining a proper asymptotic 
state, the equivalence of the longitudinal gauge-boson degrees of freedom
and the Goldstone-boson degrees of freedom contained in the fundamental Higgs
doublet becomes apparent. As such, the Sudakov logarithms for longitudinal 
gauge bosons are completely different from the ones for the transverse gauge
bosons.

Our explicit one-loop results are in agreement with the calculations
in the literature. At two-loop level, however, we do not observe a mere 
exponentiation of the one-loop results, in contrast to claims in the 
literature. The non-exponentiating terms in our two-loop
result ori\-ginate from the mass gap between the massless photon and the 
massive $Z$ boson in the neutral sector of the SM. The cancellation that 
takes place in unbroken gauge theories, leading to exponentiation, is 
upset by the fact that the on-shell poles for photons and $Z$ bosons do not 
coincide. We find that the corresponding 
non-exponentiating terms originate from the collinear domain and involve soft 
energies above the gauge-boson mass scale $M$.

{}From the explicit two-loop calculation we furthermore conclude that, 
as far as the virtual Sudakov logarithms are concerned, the SM will never 
completely behave like an unbroken theory at high energies. This is a 
consequence of the fact that the photon is strictly massless, being the gauge 
boson associated with the unbroken electromagnetic gauge group $U(1)_{\rm em}$.

A complementary study of the electroweak Sudakov logarithms for real 
collinear-soft gauge-boson emission processes is in progress.

\section*{Acknowledgements} W.B. would like to thank J.~Bl\"umlein and 
T.~Riemann for the kind invitation to a highly stimulating and enjoyable
workshop.


\begin{thebibliography}{99}

\bibitem{LCreport} E.~Accomando et.~al, Phys.~Rep.~299~(1998)~1.
\bibitem{size} P.~Ciafaloni and D.~Comelli, Phys.~Lett.~B446 (1999) 278; 
               M.~Beccaria et.~al, Phys.~Rev.~D61 (2000) 073005.
\bibitem{wwhea} W.~Beenakker et.~al, Nucl.~Phys.~B410 (1993) 245; 
                Phys.~Lett.~B317 (1993) 622.
\bibitem{sudakov} V.V.~Sudakov, Sov.~Phys.~JETP~3 (1956) 65.
\bibitem{nonexp} W.~Beenakker and A.~Werthenbach, hep-ph/0005316.
\bibitem{kuehn1} J.H.~K\"uhn and A.A.~Penin, hep-ph/9906545. 
\bibitem{paolo2} P.~Ciafaloni and D.~Comelli, Phys.~Lett.~B476 (2000) 49.
\bibitem{melles} V.S.~Fadin et.~al, Phys.~Rev.~D61 (2000) 094002.
\bibitem{taylor} J.~Frenkel and J.C.~Taylor, Nucl.~Phys.~B116 (1976) 185.
\bibitem{coulomb} W.~Beenakker and A.~Werthenbach, in preparation.
\bibitem{expon} V.G.~Gorshkov et.~al, Phys.~Lett.~22 (1966) 671; 
                J.M.~Cornwall and G.~Tiktopoulos, Phys.~Rev.~D13 (1976) 3370;
                Phys.~Rev.~D15 (1977) 2937. 
\bibitem{frenkel} J.~Frenkel and R.~Meuldermans, Phys.~Lett. B65 (1976) 64; 
                  J.~Frenkel, Phys.~Lett. B65 (1976) 383.

\end{thebibliography}
\end{document}